\documentclass{article}

\pdfoutput=1

\usepackage[margin=1in,letterpaper]{geometry}

\usepackage{amsmath,amsfonts,amssymb}
\usepackage{graphicx}
\usepackage{subfig}
\usepackage{siunitx}
\usepackage{booktabs}
\usepackage{xcolor}
\usepackage{caption}
\usepackage[colorlinks=true, allcolors=blue]{hyperref}

\usepackage{tikz}
\usetikzlibrary{arrows}
\usetikzlibrary{arrows.meta}
\tikzset{>={Latex[scale=1.25]}}
\definecolor{C0}{HTML}{1F77B4}
\definecolor{C1}{HTML}{FF7F0E}
\definecolor{C2}{HTML}{2CA02C}
\definecolor{C3}{HTML}{D62728}
\definecolor{C4}{HTML}{9467BD}
\definecolor{C5}{HTML}{8C564B}

\graphicspath{{figures/}}
 
\usepackage{authblk}
 
\title{Control Flow Graph Modifications for Improved RF-Based Processor Tracking Performance\thanks{Distribution Statement A: Approved for Public Release, Distribution Unlimited} \thanks{Preprint of: Mark Chilenski, George Cybenko, Isaac Dekine, Piyush Kumar, Gil Raz, ``Control Flow Graph Modifications for Improved RF-Based Processor Tracking Performance,'' Proc.\ SPIE 10630, Cyber Sensing 2018, 106300I (3 May 2018); DOI: 10.1117/12.2316361; \url{https://doi.org/10.1117/12.2316361}}}

\author[a]{Mark Chilenski\thanks{Send correspondence to: Mark Chilenski, mark.chilenski [\emph{at}] stresearch [\emph{dot}] com}}
\author[b]{George Cybenko}
\author[a]{Isaac Dekine}
\author[a]{Piyush Kumar}
\author[a]{Gil Raz}
\affil[a]{Systems \& Technology Research}
\affil[b]{Dartmouth College}

\date{September 19, 2018}

\begin{document} 
\maketitle

\begin{abstract}
Many dedicated embedded processors do not have memory or computational resources to coexist with traditional (host-based) security solutions. As a result, there is interest in using out-of-band analog side-channel measurements and their analyses to accurately monitor and analyze expected program execution. In this paper, we describe an approach to this problem using externally observable multi-band radio frequency (RF) measurements to make inferences about a program's execution.  Because it is very difficult to identify individual instructions solely from their RF emissions, we compare RF measurements with the constrained execution logic of the program so that multiple RF measurements over time can effectively track program execution dynamically.  In our approach, a program's execution is modeled by control flow graphs (CFG) and transitions between nodes of such graphs. We demonstrate that tracking performance can be improved through applications program modifications such as changing basic block transition properties and/or adding new basic blocks that are highly observable.  In addition to demonstrating these principled approaches on some simple programs, we present initial results on the complexity and structure of real-world applications programs, namely \texttt{gzip} and \texttt{md5sum}, in this modeling framework.
\end{abstract}

\section{Introduction}
All digital logic and data in a computing system are subject to compromise by an attacker. As a result, any security monitoring or analysis done within the same communications and computational fabric is also subject to compromise.  Moreover, many embedded processors do not have the additional memory or computational resources to provide their own security solutions. As a result, there is interest in using analog side-channel measurements and their analyses to monitor expected program execution \cite{DARPA-LADS,cybenko2018large}. We describe an approach to this problem that exploits measurements of involuntary/unintended electromagnetic emissions from commodity processors and which analyzes those measurements to estimate aspects of the embedded software execution. It is of particular importance to detect deviations from the logically constrained execution paths determined by the expected program structure. 

To detect deviations in program execution at fine granularity (e.g., changes of a single basic block) requires understanding of the embedded program's control flow graph (CFG) and the limits on our ability to correctly map the measured electromagnetic features to the currently executing block. In particular we explore the dynamic relationship between the unique classified features (or ``color'' as they map to a given code block or graph node) and the theoretical limitations on correctly identifying the traversal of the CFG. Towards this end we describe a hierarchy of CFG models in which different models have different execution tracking performance properties. We then describe approaches to modifying the program and hence the CFG such that the overall program functionality is retained with limited impact on execution time while dramatically enhancing the ability to correctly track the program and hence detect deviations from expected behavior. 

Detecting cyber agents using intrusion detection systems (IDS) running in the same machine creates new vulnerabilities and is therefore inherently risky. In such situations, cyber agents could have full access to the machine state and can  circumvent common detection checks. To address this problem, applications programs are often sandboxed within virtual machines in which a hypervisor can inspect the execution within the sandbox, thereby separating the IDS software outside the sandbox from the cyber agent execution space within the sandbox. Unfortunately, low-powered embedded devices prevalent in the Internet-of-Things (IoT) and industrial control systems (ICS) lack the computational horsepower to implement such safe guards \cite{blackman1999design}.

\begin{figure} 
	\centering
    \includegraphics[scale=0.5]{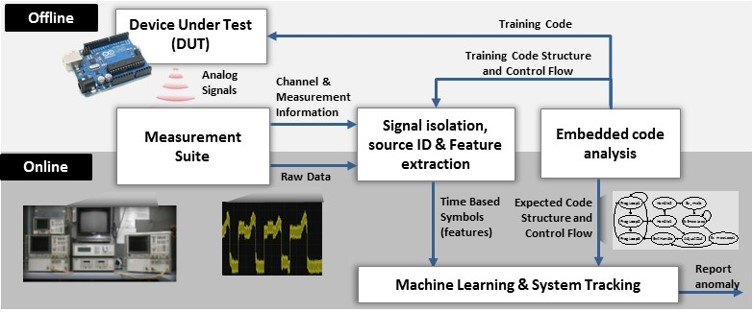}
	\caption{The {\it Leveraging Anomalous System Signals to Enable Security} (LASSES) system detects deviations in embedded program operation across an air-gap using involuntary EM analog emissions from processors. It combines: comprehensive measurements to understand the analog channel, signal filtering and isolation and feature generation for use in training and in real-time operation, code analysis functions, and program tracking and machine learning functions for detecting deviations from expected DUT operation.}
\label{fig:lassesblock}
\end{figure}

The authors and their respective organizations are jointly conducting research into methods for providing cyber defenses through analysis of involuntary analog emissions. Our method  provides a principled approach to analyze and exploit all available information sources and inform about achievable performance bounds and  quantify analog side-channel information. We analyze individual and combined contributions of analog emissions and propagation physics, signal and interference levels and waveforms, and code structure information. Figure~\ref{fig:lassesblock} depicts a notional high-level description of our real-time system.

This approach consists of three key areas of innovations:
\begin{itemize}
\item Measurements suite that consists of instantaneously broad band high fidelity measurements, as well as using cost efficient software defined radios (SDRs) and multi-modality collections. These provide comprehensive depth and breadth of data collected under multiple conditions, providing accurate waveforms and distribution of signals-of-interest and interference. Signal quality as measured by signal-to-noise-ratio (SNR) and signal-to-interference-and-noise (SINR) can be be compared against the theoretic bounds provided by an analog link analysis parameterized by emission source, distance, modality, and channel conditions. 
\item Novel data adaptive signal processing methods that utilize annotated test vectors to exploit side information about measurements, devices, and embedded code to generate a stream of (probability weighted) features related to internal device state and operations. In addition, this can find optimal bands of operation and filters to separate signals from interference and noise while retaining classification information. Our approach quantifies available information at the raw data level and the incremental information provided by the side-information using pre- and post-processing SNR and SINR.
\item Machine learning algorithms using multiple resolution dynamic models that first learn models that link known program execution to observed measurements. State tracking over time is used to validate the observed device under test (DUT) measurements fit the learned model; thereby validating the DUT is operating as expected. This state-modeling framework jointly captures program execution at (1) sub basic-block level of linear instruction execution, (2) control flow graph level modeling of transitions between basic blocks, and (3) context level modeling processing scheduling, threading, and interrupts.
\end{itemize}

By focusing primarily on RF/EM emissions, our  approach isolates the security functionality by an ``air gap'' from the target host. A major challenge that must be overcome, however, is obtaining sufficient state information via unintended emissions to accurately characterize the threat.

In this paper, we describe a modeling and analysis approach that allows us to model programs through their control flow graphs so that the RF measurements can be regarded as noisy observations of the basic blocks executing within the program.  Section~\ref{sec:modelFramework} outlines the modeling framework, Sec.~\ref{sec:expts} describes experiments and analyses we have performed on notional and real-world programs and finally Sec.~\ref{sec:summary} is a summary with discussion of future work we intend to conduct in this general problem space.

\section{Model Framework}
\label{sec:modelFramework}

In this section we describe in detail the model framework that captures program execution at the basic-block level. This allows us to address the control flow graph tracking problem that must be solved before confronting the issue of cyber intrusion detection using sensor measurements of RF emissions. Then, we outline a taxonomy of ``observability classes" of CFGs. We find that even though in many cases the CFG is not ``observable"
(in a technical sense that is explained later), the CFG can be modified in appropriate ways so as to significantly improve its tracking performance with no effect on program functionality and minimal effect on real-time and memory constraints.  

\subsection{Colored graphs and the hidden Markov model}
\label{sec:CommModel}
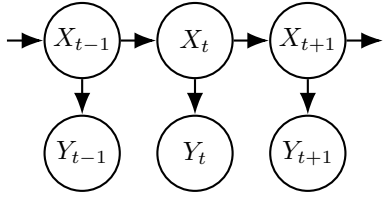
\begin{figure}
	\centering
	\begin{tikzpicture}
    	\node [draw, circle, inner sep=0, minimum size=1cm, thick, name=xtm1] at (0,0) {$X_{t-1}$};
        \node [draw, circle, inner sep=0, minimum size=1cm, thick, name=xt] at (1.5,0) {$X_t$};
        \node [draw, circle, inner sep=0, minimum size=1cm, thick, name=xtp1] at (3,0) {$X_{t+1}$};
        \node [draw, circle, inner sep=0, minimum size=1cm, thick, name=ytm1] at (0,-1.5) {$Y_{t-1}$};
        \node [draw, circle, inner sep=0, minimum size=1cm, thick, name=yt] at (1.5,-1.5) {$Y_t$};
        \node [draw, circle, inner sep=0, minimum size=1cm, thick, name=ytp1] at (3,-1.5) {$Y_{t+1}$};
        \draw[->, thick] (xtm1) -- (xt);
        \draw[->, thick] (xt) -- (xtp1);
        \draw[->, thick] (xtm1) -- (ytm1);
        \draw[->, thick] (xt) -- (yt);
        \draw[->, thick] (xtp1) -- (ytp1);
        \draw[->, thick] (-1.0,0) -- (xtm1);
        \draw[->, thick] (xtp1) -- (4.0,0);
    \end{tikzpicture}
    \caption{Graphical model of the HMM. The sequence of states $X_t$ cannot be measured directly. Instead, the sequence of symbols $Y_t$ is measured. Each state depends only on the previous state and each symbol depends only on the present state.}
    \label{fig:hmm}
\end{figure}

Software execution is modeled as a series of transitions between states on a control flow graph (CFG), where the nodes can be modules, functions, or basic blocks (depending on the desired level of granularity) and the edges are the allowed transitions between states. While each node is executing, a particular pattern of RF emissions occurs.
The sequence of states $X_t$ cannot be observed directly, instead the states must be inferred from the observed sequence of RF emissions (``symbols'') $Y_t$.
A natural formalism describing such a situation is the hidden Markov model (HMM) shown in Fig.~\ref{fig:hmm} \cite{levinson1983introduction,rabiner1989tutorial}.

It is important to note that the concept of ``state''  as used in systems theory does not correspond necessarily to a basic block in a program's CFG because the true state of a program includes the memory values as well as the program counter and the reader is encouraged to understand this difference.  We address the question of how closely basic blocks approximate system state in the more formal sense in Sec.~\ref{sec:hmmOrderEst} when we empirically investigate some real-world programs.

In this model, the sequence of hidden states $X_t$ forms a (first order) Markov chain:
\begin{gather}
	P(X_t | X_1,\dots, X_{t-1}) = P(X_t|X_{t-1})\label{eq:firstOrder}
\end{gather}
with transition matrix
\begin{gather}
	P_{ij} := P(X_t = j | X_{t-1} = i).
\end{gather}
The symbol $Y_t$ emitted at any point depends only on the current state $X_t$:
\begin{gather}
	P(Y_t|X_1,\dots, X_{t}, Y_1,\dots, Y_{t-1}) = P(Y_t|X_t).
\end{gather}
In the case of discrete symbols, the emission matrix is defined as
\begin{gather}
	o_{i\alpha} := P(Y_t=\alpha|X_t=i).
\end{gather}
While much of the literature on HMMs focuses on discrete scalar symbols $Y_t$, this formalism can handle the more general case where $Y_t$ is a multidimensional, continuous random variable simply by replacing $P(Y_t|X_t)$ with the relevant joint probability density function (PDF).
Furthermore, it is often convenient to assign a discrete scalar symbol to a general measurement using a classification or clustering algorithm.

\begin{figure}
	\centering
    \subfloat[Emission matrix]{\includegraphics[scale=0.55]{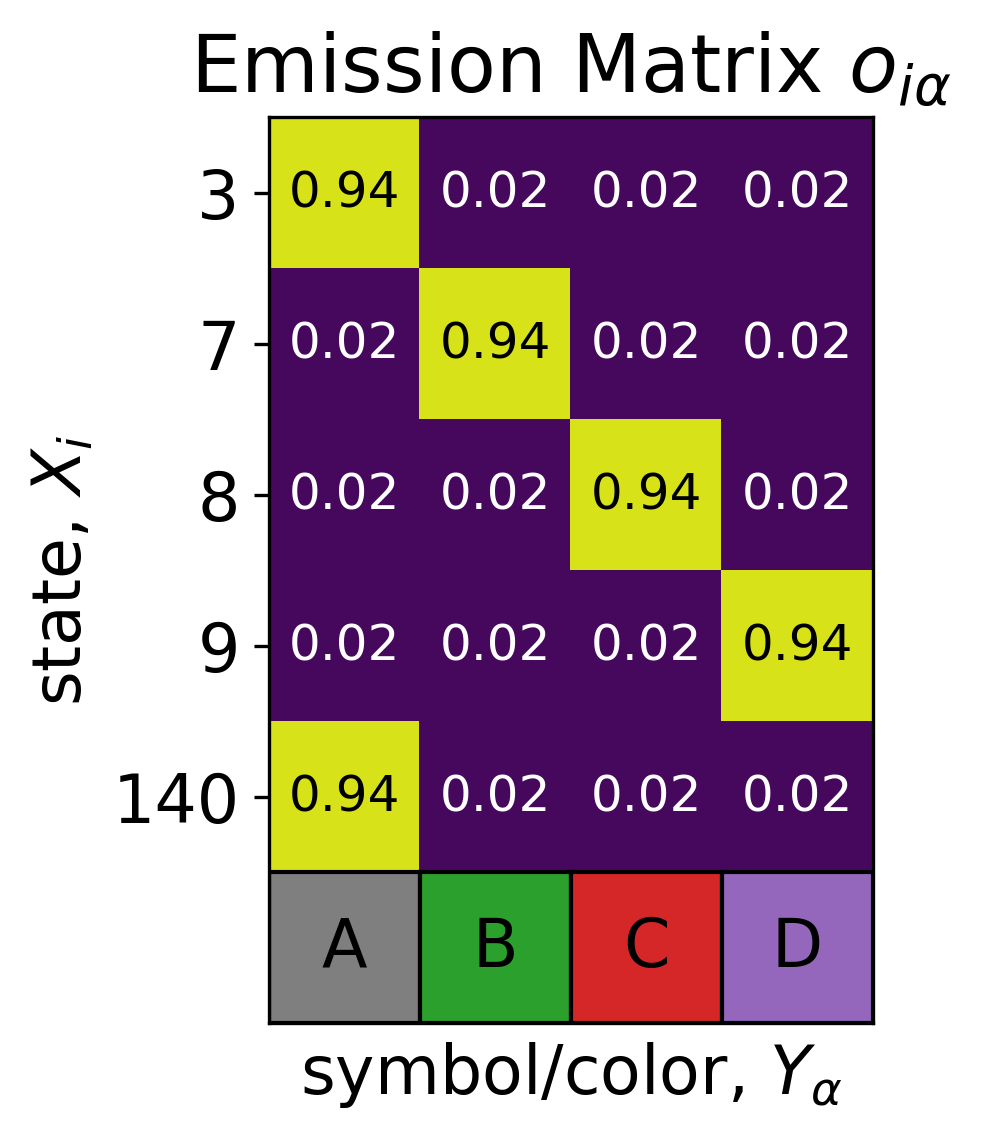}\label{sf:emis}}
    \subfloat[Colored graph]{\includegraphics[scale=0.475]{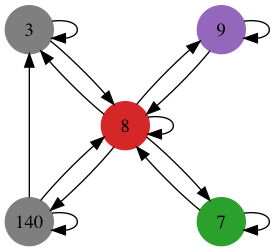}\label{sf:cfg}}
    
    \subfloat[Featuregram with associated colors]{\includegraphics[scale=0.425]{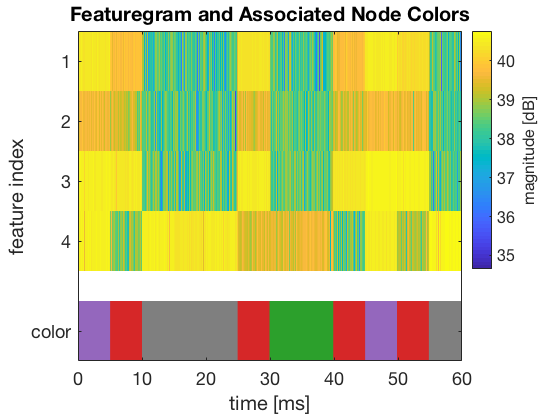}\label{sf:fg}}
	\caption{An emission matrix \protect\subref{sf:emis} which has only one highly-probable symbol per state can be described as assigning a ``color'' to each state on the CFG \protect\subref{sf:cfg}. The featuregram \protect\subref{sf:fg} shows the four-dimensional, continuous-valued feature vector and the corresponding colors assigned by a classifier.}
	\label{fig:coloredGraph}
\end{figure}

As a further abstraction, the discrete symbols may be taken to correspond to ``colors'' emitted by each state.
If the emission distribution preferentially emits a single color for each state, the model may be described as a node-colored directed graph. An example of this is shown in Fig.~\ref{fig:coloredGraph}.
When multiple colors may be emitted by a state, the graph is said to be multi-colored, and can be reduced to a simple node-colored directed graph by splitting each multi-colored node.
As will be seen, node-colored directed graphs provide a useful framework for reasoning about the properties of control flow graphs.

\subsection{Observability classes of colored graph models}
\label{sec:taxonomy}

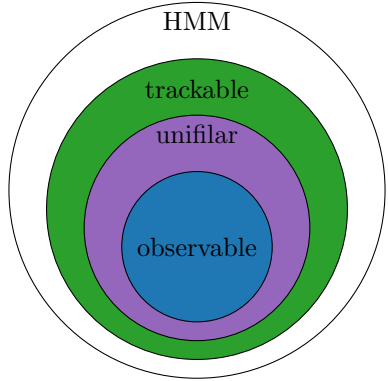
\begin{figure}
	\centering
	\begin{tikzpicture}
    	\draw (0,0.75) circle (2.5cm);
        \draw[fill=C2] (0,0.5) circle (2cm);
        \draw[fill=C4] (0,0.25) circle (1.5cm);
        \draw[fill=C0] (0,0) circle (1cm);
        \node at (0,0) {observable};
        \node at (0,1.5) {unifilar};
        \node at (0,2.1) {trackable};
        \node at (0,3) {HMM};
    \end{tikzpicture}
    \caption{Hierarchy of colored graph models. The hidden Markov model (HMM) is the most general case, for which no guarantees about the ability to infer the hidden states can be made. Models from deeper in the hierarchy provide better guarantees on tracking performance.}
    \label{fig:graphHierarchy}
\end{figure}

As mentioned earlier, a node-colored directed graph (corresponding to a CFG) can be identified by a state model in which the nodes of the graph represent states and the edges represent allowed transition between states. The most general description of such a state model is provided by an HMM. The structure of the underlying colored graph may not permit unique reconstruction of the state sequence in general. There are, however, restricted classes of models that enable more accurate tracking in various use cases:
\begin{itemize}
	\item In a \emph{trackable} model the number of hypotheses consistent with an observation sequence grows polynomially in the length of the observation sequence \cite{crespi2008theory}.
    \item In a \emph{unifilar} model the next state $X_{t+1}$ is uniquely determined given the current state $X_t$ and the next symbol $Y_{t+1}$ \cite{sheng2005}.
    \item In an \emph{observable} model the state $X_t$ can be uniquely determined given the observation sequence $Y_1,\dots,Y_t$, after some fixed initial burn-in \cite{jungers2011observable}.
\end{itemize}
This hierarchy of graph classes is shown in Fig.~\ref{fig:graphHierarchy}.

\subsection{Graph modification to improve observability}
In principle, it is possible to change the observability class of a graph by adding or deleting states, by adding or deleting transitions between nodes, or both. Given that the specific way in which a control flow graph is connected as well as the transition probabilities between different nodes determines software functionality, removing states or modifying transition probabilities significantly is not guaranteed to preserve program function.

On the other hand, the addition of states (nodes) constructed to have no effect on program function (e.g., a short loop of no-op instructions), but which emit a distinct color, can be very useful. Inserting such ``indicator blocks'' at strategic points in a program can change the observability class and improve tracking performance, which we demonstrate in Sec.~\ref{sec:indicatorExperiment}.

It is important to remark that such program modifications, while not changing program semantics, could change a program's memory footprint and/or its real-time performance characteristics.  Such modifications must accordingly be done with such constraints in mind.

\section{Experiments and Results}
\label{sec:expts}
In this section, we describe the experiments conducted on test programs and real software programs and the results obtained. In particular, in Sec.~\ref{sec:indicatorExperiment} we describe the experimental setup and data collection procedure for simple test programs, and then demonstrate the efficacy of our CFG modification approach. We show that impressive results for CFG tracking can be obtained for such test programs. 

Carrying out a similar analysis for real software programs is considerably more difficult, however. For example, in many cases the source code may not be available and, even if it is, low-level program structure may be significantly altered by an optimizing compiler so that information on the program's structure may be difficult to obtain. Therefore tools must be used to extract the control flow graph of the program as it will be executed.

Many dynamic profiling and tracing tools also have limited support on embedded platforms. Also, as we will see later, the time scale for transitions between different basic blocks of realistic CFGs is typically much smaller than that of the observation model corresponding to RF emissions. Hence ``coarse-graining" techniques must be applied to the original CFG to extract a coarsened version that corresponds accurately to the observation model. In Sec.~\ref{sec:realsoftware} we show some results related to the extraction of control flow graphs and their properties for some simple realistic software programs -- \texttt{gzip} and \texttt{md5sum}. We also find the appropriate model framework to track execution of these programs. In the near future, we intend to apply our CFG modification and tracking techniques to real software programs such as these and also study use cases for cyber intrusion detection as described in Sec.~\ref{sec:summary}.

\subsection{Experimental demonstration of pathology mitigation}
\label{sec:indicatorExperiment}
Our first experiment utilized the Arduino Uno, a common Internet-of-Things (IoT) device which exemplifies the security challenges and motivations. The Uno, with a \SI{16}{MHz} processor clock, \SI{32}{kB} of program flash memory, and no tractable ability to execute parallel execution threads, is unequipped to perform any host-based security activities. Furthermore, the performance capabilities of this board and processor are very similar to millions of deployed hardware devices.

\begin{figure}
	\centering
    \subfloat[Transition matrix]{\includegraphics[scale=0.6]{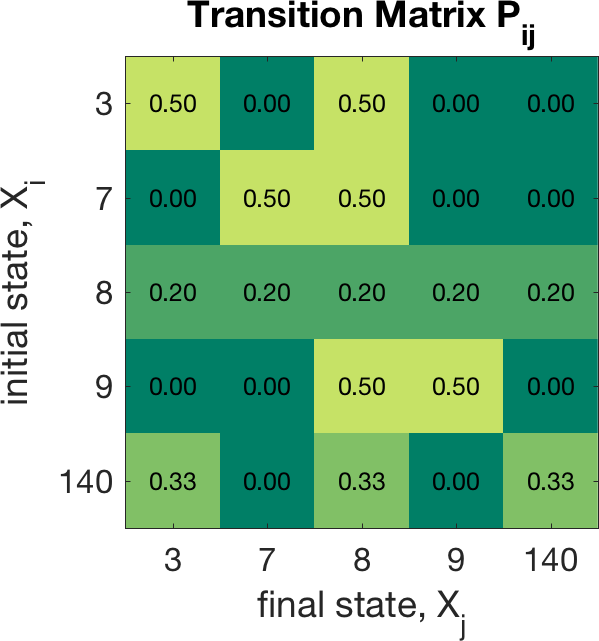}\label{sf:equalPij}}
    \hspace{3.5em}
    \subfloat[Emission matrix]{\hspace{-3em}\includegraphics[scale=0.7]{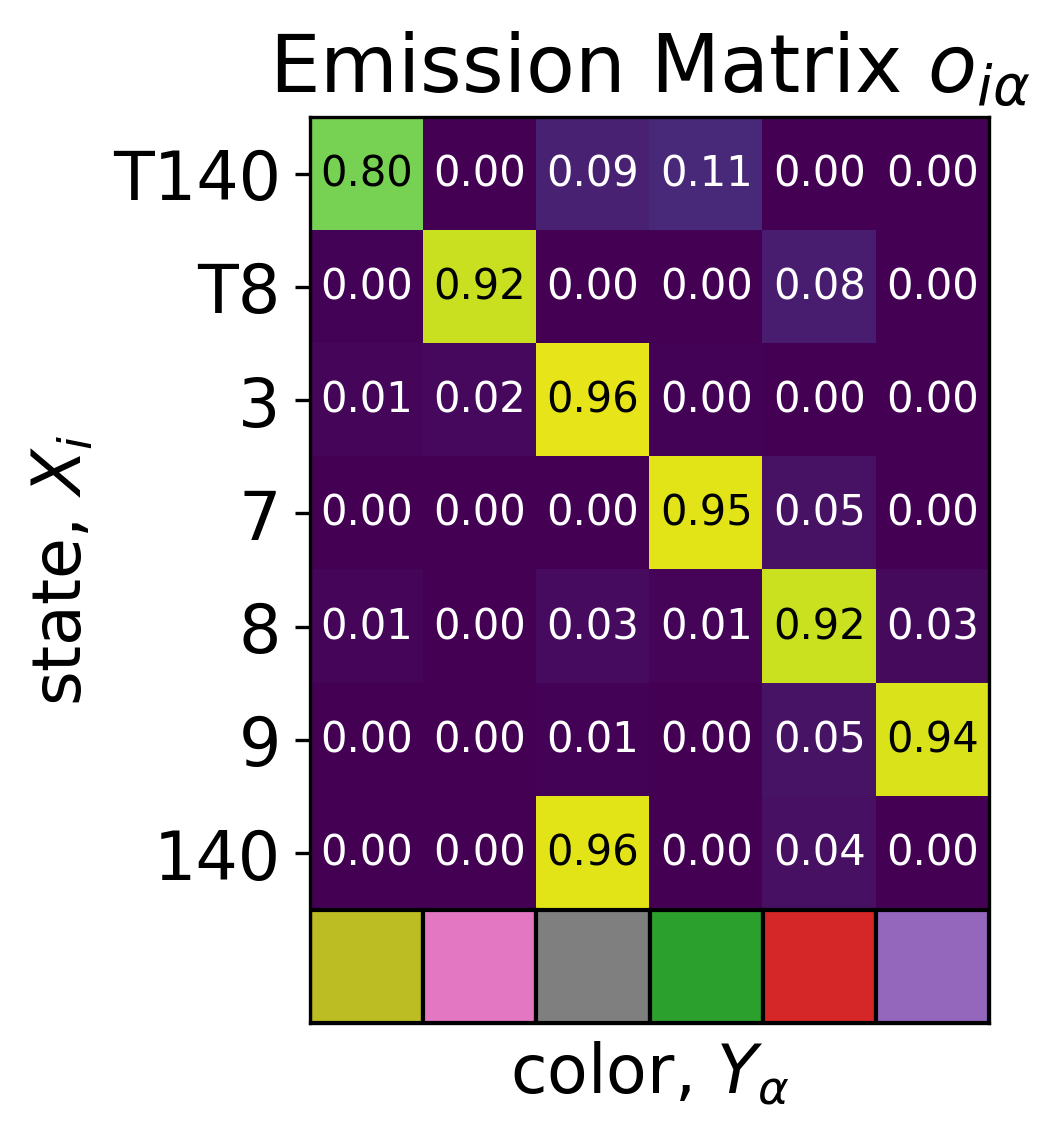}\label{fig:emisExpt}}
    \caption{Hidden Markov model for the Arduino tracking experiment: \protect\subref{sf:equalPij} (unmodified) transition matrix and \protect\subref{fig:emisExpt} observed emission matrix. The emission matrix is computed after applying the classifier but before binning into \SI{5}{ms} steps.}
\end{figure}

This first experiment, designed to facilitate the development of control flow graph tracking algorithms and pathology mitigation, was composed as follows:
\begin{itemize}
\item	create basic blocks, $b_1,\dots,b_n$, where each basic block is a loop that performs a unique set of instructions as shown in Fig.~\ref{fig:basicblock};
\begin{figure}
	\centering
    \includegraphics[scale=0.55]{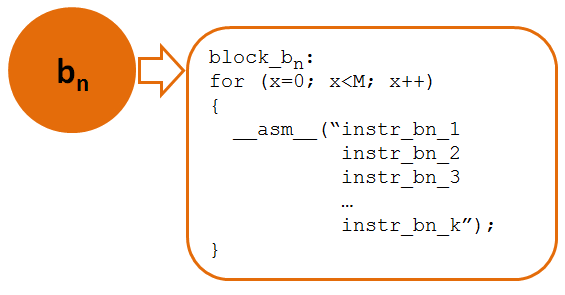}
	\caption{Example basic block.}
    \label{fig:basicblock}
\end{figure}
\item	individually execute and perform RF measurements on each basic block, using the measurements to train a classifier;
\item	create a control flow graph that utilizes a subset of these basic blocks;
\item	perform random walks through this control flow graph, such that at each time step, the next basic block to execute will be randomly selected from the available legal transitions from the presently executing block, with probabilities shown in Fig.~\ref{sf:equalPij}.
\end{itemize}

\begin{figure}[t!]
	\centering
    \subfloat[Original: unobservable]{\includegraphics[scale=0.55]{GarduinoExpt}\label{sf:orig}}
    \hspace{4em}
    \subfloat[Indicator blocks added: unifilar]{\includegraphics[scale=0.55]{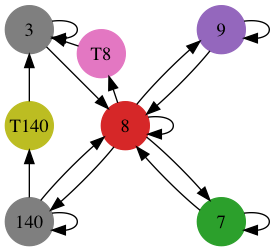}\label{sf:mit}}
	\caption{Control flow graphs for the Arduino tracking experiment: \protect\subref{sf:orig} original and \protect\subref{sf:mit} with indicator blocks added. The modified CFG is unifilar and not observable because of the self-transitions at nodes 3 and 140 which constitute two separated cycles with the same coloring.}
    \label{fig:arduinoCFGs}
\end{figure}

We started with the unobservable CFG in Fig.~\ref{sf:orig} and the transition matrix in Fig.~\ref{sf:equalPij}, where the blocks are composed of loops of between 20 and 105 instructions.
We set the number of loop executions in each block such that each ``step'' through a block lasted \SI{5}{ms}.
We then added indicator blocks to produce the unifilar CFG in Fig.~\ref{sf:mit}.

\begin{figure}
	\centering
    \includegraphics[scale=0.3]{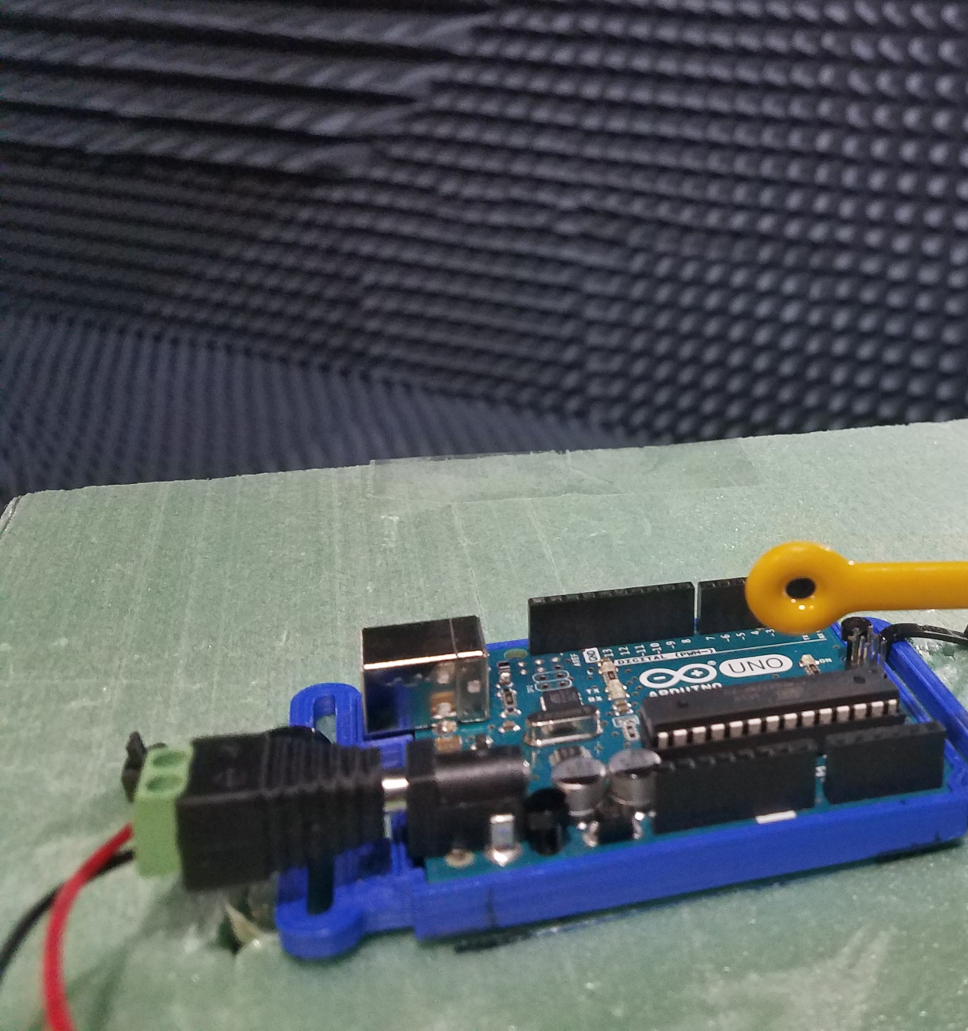} 
    \caption{Measurement setup: Arduino Uno R3, near field magnetic probe \SI{10}{mm} above the device, inside an anechoic RF chamber}
    \label{fig:arduinoNFPMeas}
\end{figure}

The RF emissions were measured at a range of \SI{10}{mm}, using a near field probe sensitive to the magnetic field from \SI{300}{kHz} to \SI{1}{GHz}. 
This probe was sampled at \SI{4}{GHz} using a Texas Instruments ADC12J4000 analog-to-digital conversion system.
The measurement setup is shown in Fig.~\ref{fig:arduinoNFPMeas}.
We computed the short-time Fourier transform using a \SI{66}{\micro s} window length, applied a random forest classifier to each time point in the resulting spectrogram, then binned the results to a \SI{5}{ms} time resolution by taking the mode in each bin.
The emission matrix from the classifier (but prior to binning) is shown in Fig.~\ref{fig:emisExpt}.
The effect of binning is to make all of the states except T140 single-colored.
The test with indicator blocks added used the same path through the CFG as the test without indicator blocks -- the indicator blocks were simply added at the appropriate transitions.

\begin{figure}
	\centering
    \subfloat[Original]{\includegraphics[width=0.45\textwidth]{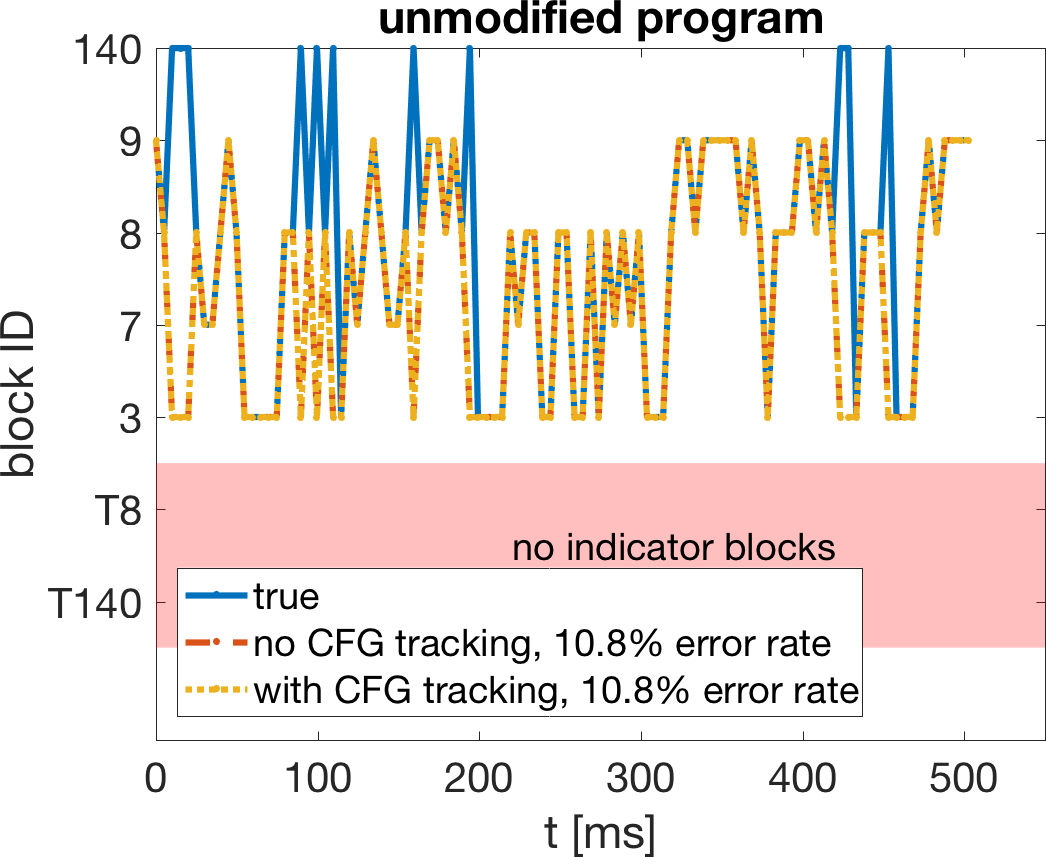}\label{sf:trackOrig}}
    \hspace{2em}
    \subfloat[Tracking blocks added]{\includegraphics[width=0.45\textwidth]{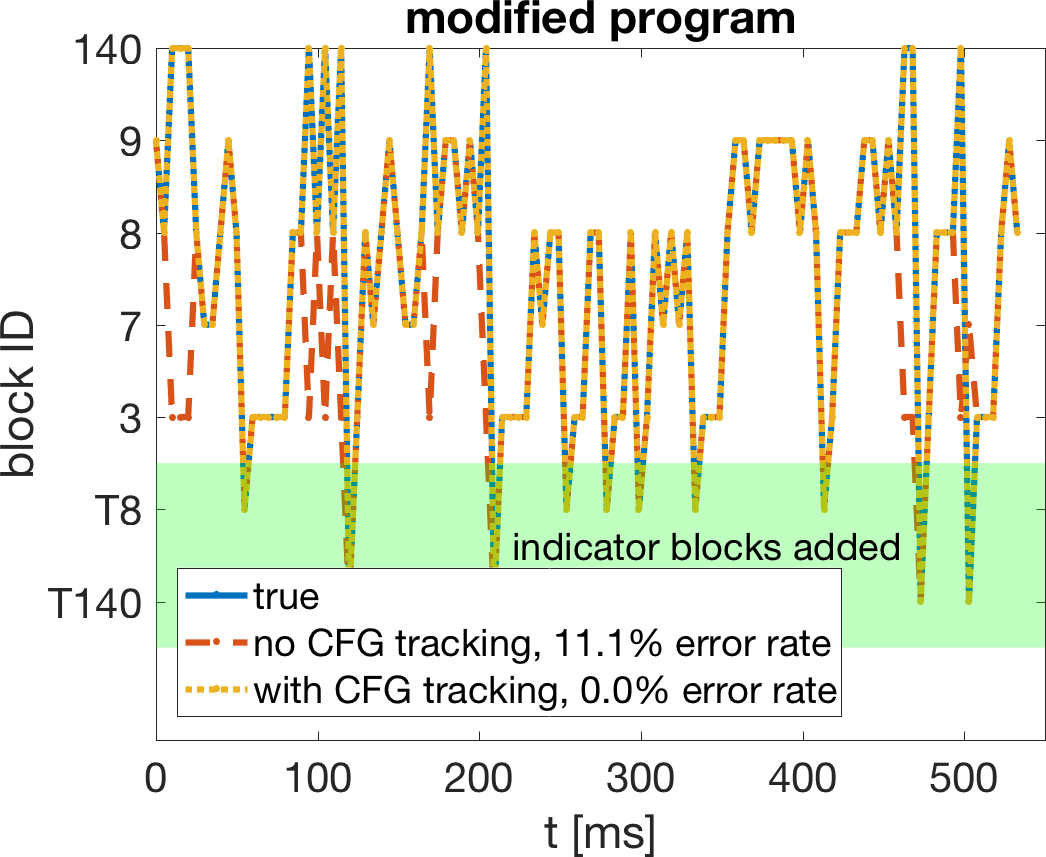}\label{sf:trackMit}}
    \caption{Tracking performance for the \protect\subref{sf:trackOrig} original and \protect\subref{sf:trackMit} modified CFGs. The solid blue line is the true path, the dashed orange line is the result of classifying each moment in time (i.e., \emph{not} applying the Viterbi algorithm), and the dotted yellow line is the result of applying the Viterbi algorithm to the entire symbol sequence. In the original CFG, applying the Viterbi algorithm does not improve on the 10.8\% error rate. In the CFG with tracking blocks added, the Viterbi algorithm provides perfect tracking performance, as is expected for a unifilar graph.}
    \label{fig:trackingPerformance}
\end{figure}

The results are shown in Fig.~\ref{fig:trackingPerformance}. In the figure, the ``no CFG tracking'' curve is the state inferred directly by the classifier with no knowledge of the CFG and the ``with CFG tracking'' curve was constructed by applying the Viterbi algorithm to find the most likely path given knowledge of the transition matrix, the classifier confusion matrix, and the entire symbol sequence. Without indicator blocks, CFG tracking has no effect, and the error rate (number of blocks wrong divided by total number of blocks) is 10.8\%. With indicator blocks, CFG tracking obtains perfect reconstruction of the sequence of states, as is expected for a unifilar graph.

\subsection{CFG extraction and model framework for real software}\label{sec:realsoftware}
In order to assess the prognosis for basic block-level tracking in real software, we have begun collecting instruction-level execution traces from common Linux tools using GDB \cite{gdb}.
Presented here are the results from the \texttt{gzip} and \texttt{md5sum} programs on Ubuntu running on x86\_64 applied to 100 randomly-generated \SI{1}{kB} files.
The instruction-level traces are post-processed into basic blocks based on jump, call, and return instructions.
The CFG is then extracted based on which transitions are observed in the 100 runs -- no static analysis of possible jump targets has been performed.
The resulting CFG has many blocks with in-degree and/or out-degree of one.
To make the subsequent analysis tractable, whenever a block's in-degree and its parent's out-degree are both one, we merge the block and its parent into a ``reduced block.''

\subsubsection{Blocks, reduced blocks have few instructions}
\begin{figure}
	\centering
    \subfloat[\texttt{gzip}]{\includegraphics[width=0.45\textwidth]{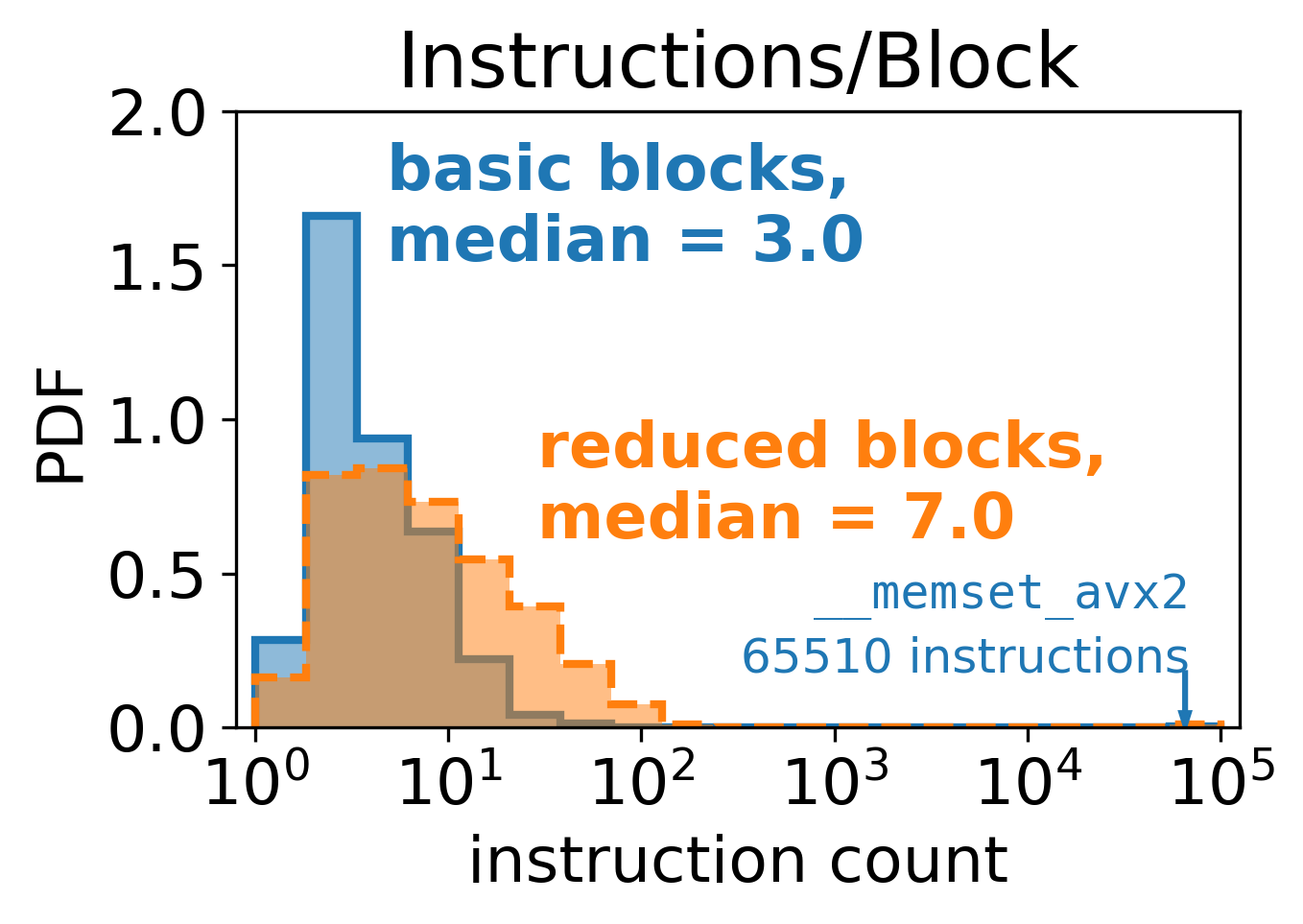}\label{sf:gzipInstDist}}
    \subfloat[\texttt{md5sum}]{\includegraphics[width=0.45\textwidth]{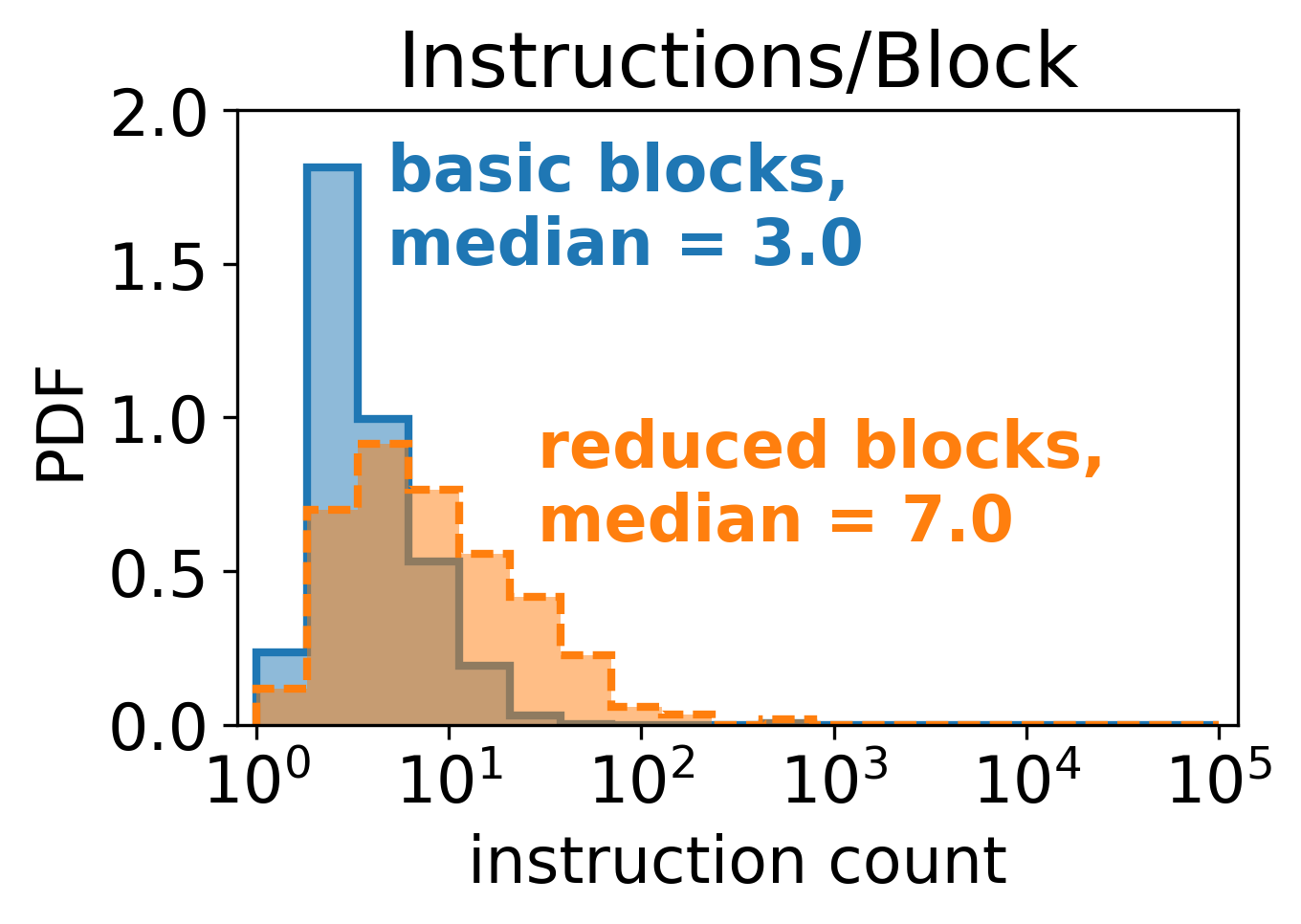}\label{sf:md5sumInstDist}}
    \caption{Distribution of the number of instructions per basic (blue, solid border) and reduced (orange, dashed border) block. Note the logarithmic scale -- the probability density function (PDF) on the vertical scale is that of $\log_{10}(\text{instruction count})$. Note that \texttt{gzip} has a \emph{basic} block with \num{65510} instructions. This is part of \texttt{\_\_memset\_avx2}, and is initializing a large chunk of memory.}
    \label{fig:instHist}
\end{figure}

The distribution of the number of instructions per block is shown in Fig.~\ref{fig:instHist}.
In both programs, the median number of instructions per basic block is about 3, which is comparable to (if not slightly lower than) what was reported in \cite{patel2000increasing}.
Working in terms of reduced blocks increases the median number of instructions to 7 and the right-hand tail of the distribution is pulled out to include reduced blocks with a few hundred instructions.
The shortness of basic and reduced blocks indicates that time resolution will be a key obstacle to overcome to obtain block-level tracking of real software -- 7 instructions is about \SI{0.4}{\micro s} on an Arduino Uno, and will be even shorter on more powerful platforms.

\subsubsection{Control Flow Graphs have very sparse transition matrices}
\begin{figure}
	\centering
    \subfloat[\texttt{gzip}]{\includegraphics[width=0.45\textwidth]{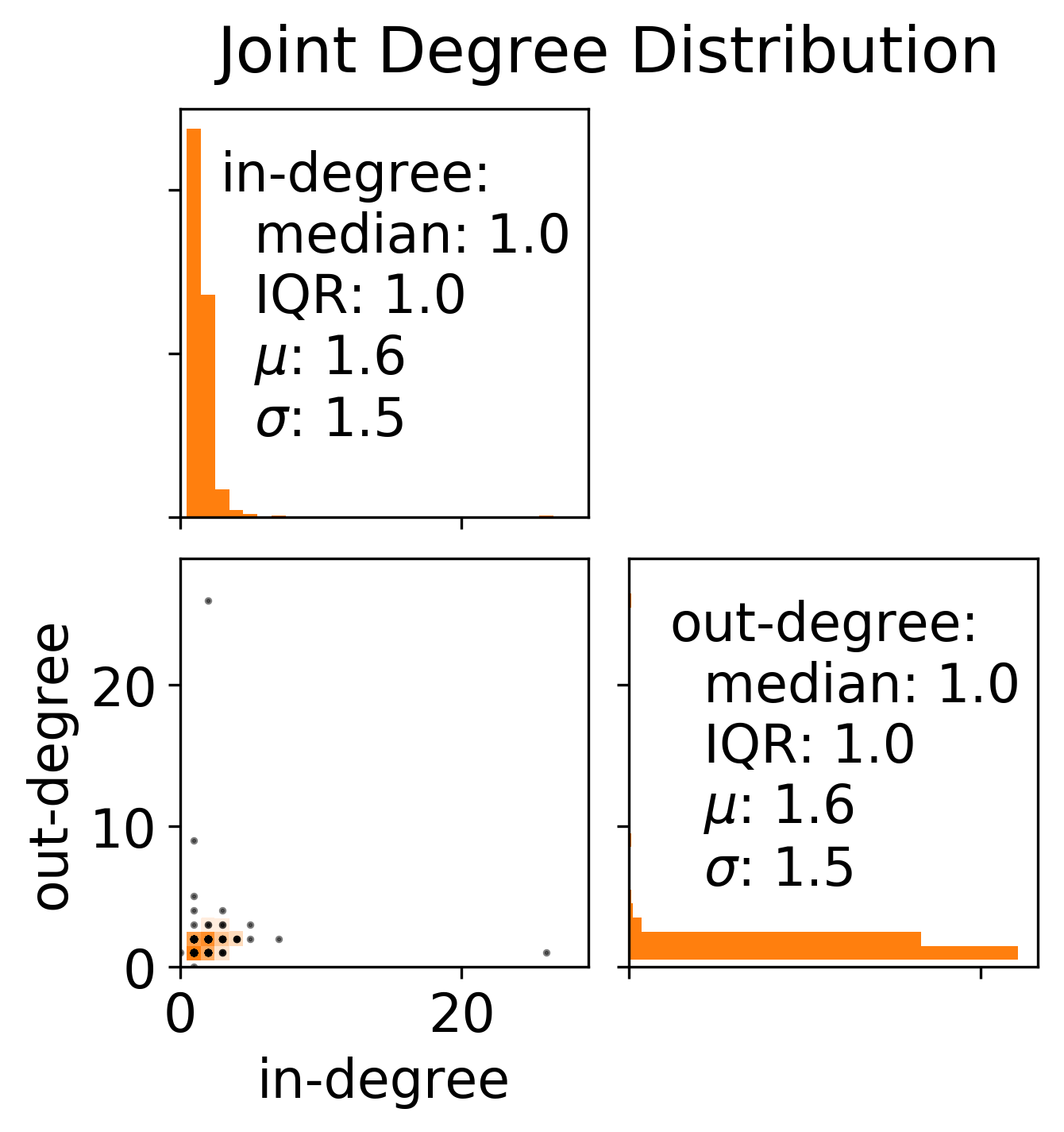}\label{sf:gzipDegDist}}
    \subfloat[\texttt{md5sum}]{\includegraphics[width=0.45\textwidth]{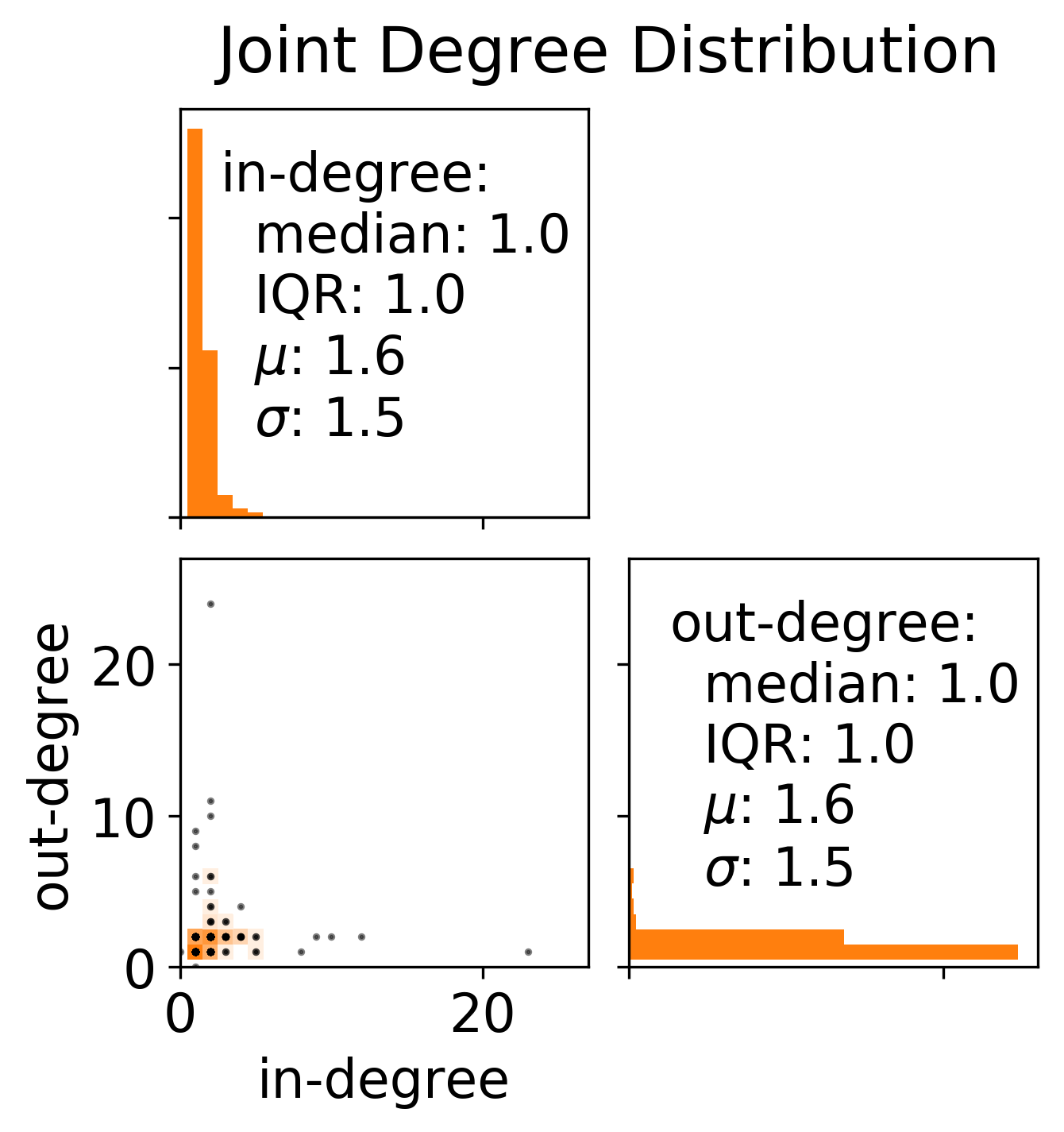}\label{sf:md5sumDegDist}}
    \caption{Joint (lower left) and marginal (diagonal) distributions of in- and out-degree of the reduced block CFGs. In the joint distribution, the shading is proportional to logarithm of the counts in the bin, and the black points indicate which bins are populated -- the distributions are both \emph{very} strongly concentrated at $\text{in-degree}=\text{out-degree}=1$. Also of note is that some nodes have high in-degree or high out-degree, but none have both -- the tails of the joint distribution form an ``L'' shape.}
    \label{fig:degDist}
\end{figure}

The joint distribution of in- and out-degree for the reduced block CFG is shown in Fig.~\ref{fig:degDist}, and some properties of the basic and reduced CFGs are given in Tab.~\ref{tab:blockCounts}.
Even in the reduced CFG, there are still many nodes with $\text{in-degree}=\text{out-degree}=1$; as suggested by the fact that the number of edges in the reduced CFG is $\sim\! 1.6\times$ the number of reduced blocks, the transition matrices are \emph{very} sparse.
This improves the prognosis for observability, because a lower average out-degree means there are fewer opportunities for the conditions for observability, unifilarity, or trackability to be violated.

\begin{table}
	\centering
    \caption{Properties of the Full and Reduced CFGs}
    \label{tab:blockCounts}
    \begin{tabular}{lS[table-format=4.0]S[table-format=4.0]S[table-format=3.0]S[table-format=3.0]}
    	\toprule
        Program & \multicolumn{2}{c}{Full CFG} & \multicolumn{2}{c}{Reduced CFG}\\
        \cmidrule(lr){2-3}\cmidrule(l){4-5}
         & {Blocks} & {Edges} & {Blocks} & {Edges}\\
        \midrule
        \texttt{gzip} & 936 & 1134 & 348 & 546\\
        \texttt{md5sum} & 1446 & 1700 & 457 & 711\\
        \bottomrule
    \end{tabular}
\end{table}

\subsubsection{A first-order Markov chain best describes the datasets}
\label{sec:hmmOrderEst}
When the HMM was introduced in Sec.~\ref{sec:CommModel}, Eq.~(\ref{eq:firstOrder}) defined a first-order Markov chain: the next state only depends on the current state.
However, computer programs have memory, and the next state (basic block) may depend on several of the previous states (basic blocks) as well as the memory state of the program.
A Markov chain with memory (or order) $m$ has the transition distribution
\begin{gather}
	P(X_t|X_1,\dots,X_{t-1}) = P(X_t|X_{t-m},\dots,X_{t-1}).
\end{gather}
Such a model has $|S|^m(|S|-1)$ parameters, where $|S|$ is the number of states.
Table~\ref{tab:numparam} gives the number of states, number of observations, and number of parameters for the \texttt{gzip} and \texttt{md5sum} datasets.
Because of the exponential growth in the number of parameters, the $m=2$ case has substantially more parameters than observations.
Therefore, it is reasonable to expect that the data will only support $m\leq 1$.

\begin{table}
	\centering
    \caption{Number of Parameters for Various Markov Chain Orders}
    \label{tab:numparam}
    \begin{tabular}{lS[table-format=3.0]S[table-format=1.1e1]S[table-format=1.1e1]S[table-format=1.1e1]}
    	\toprule
        Program & {States} & {Observations} & \multicolumn{2}{c}{Parameters, $n_{\theta_m}$}\\
        \cmidrule(l){4-5}
         & {$|S|$} & {$n_D$} & {$m=1$} & {$m=2$}\\
         \midrule
         \texttt{gzip} & 348 & 5.0e6 & 1.2e5 & 4.2e7\\
         \texttt{md5sum} & 457 & 2.2e6 & 2.1e5 & 9.5e7\\
         \bottomrule
    \end{tabular}
\end{table}

This assertion can be made rigorous using model selection tools, including \cite{singer2014}:
\begin{itemize}
	\item Akaike information criterion:
    \begin{gather}
    	AIC_m = 2n_{\theta_m} - 2\ln L_m,\label{eq:AIC}
    \end{gather}
    where $n_{\theta_m}$ is the number of parameters for the model with memory $m$ and $L_m=\max_{\theta_m} P(D|\theta_m,m)$ is the probability of the data $D$ under the maximum likelihood estimate. A lower $AIC_m$ indicates a better model.
    \item Bayesian information criterion:
    \begin{gather}
    	BIC_m = n_{\theta_m}\ln n_D - 2\ln L_m,\label{eq:BIC}
    \end{gather}
    where $n_D$ is the number of observations. A lower $BIC_m$ indicates a better model.
    \item Model evidence (also known as marginal likelihood):
    \begin{gather}
    	Z_m = P(D|m) = \int P(D|\theta_m,m)P(\theta_m|m)\,\mathrm{d}\theta_m,
    \end{gather}
    which is the probability of the data averaged over the possible values of the parameters, and $P(\theta_m|m)$ is the prior distribution representing any prior knowledge about the parameters $\theta_m$. In this application, we use a Dirichlet distribution with $\alpha_{ij}=1$ as our prior distribution when computing $Z_m$. A higher $Z_m$ indicates a better model.
\end{itemize}
In the limit of $n_D\gg n_{\theta_m}$ \cite{schwarz1978},
\begin{gather}
	BIC_m \approx -2\ln Z_m,\label{eq:BICLimit}
\end{gather}
so we report $-2\ln Z_m$ so that all quantities are on the same scale.
All three of these metrics implement a tradeoff between the model complexity (the number of parameters relative to the number of observations) and the how well the model fits the data.

\begin{figure}
	\centering
    \subfloat[\texttt{gzip}]{\includegraphics[width=0.4\textwidth]{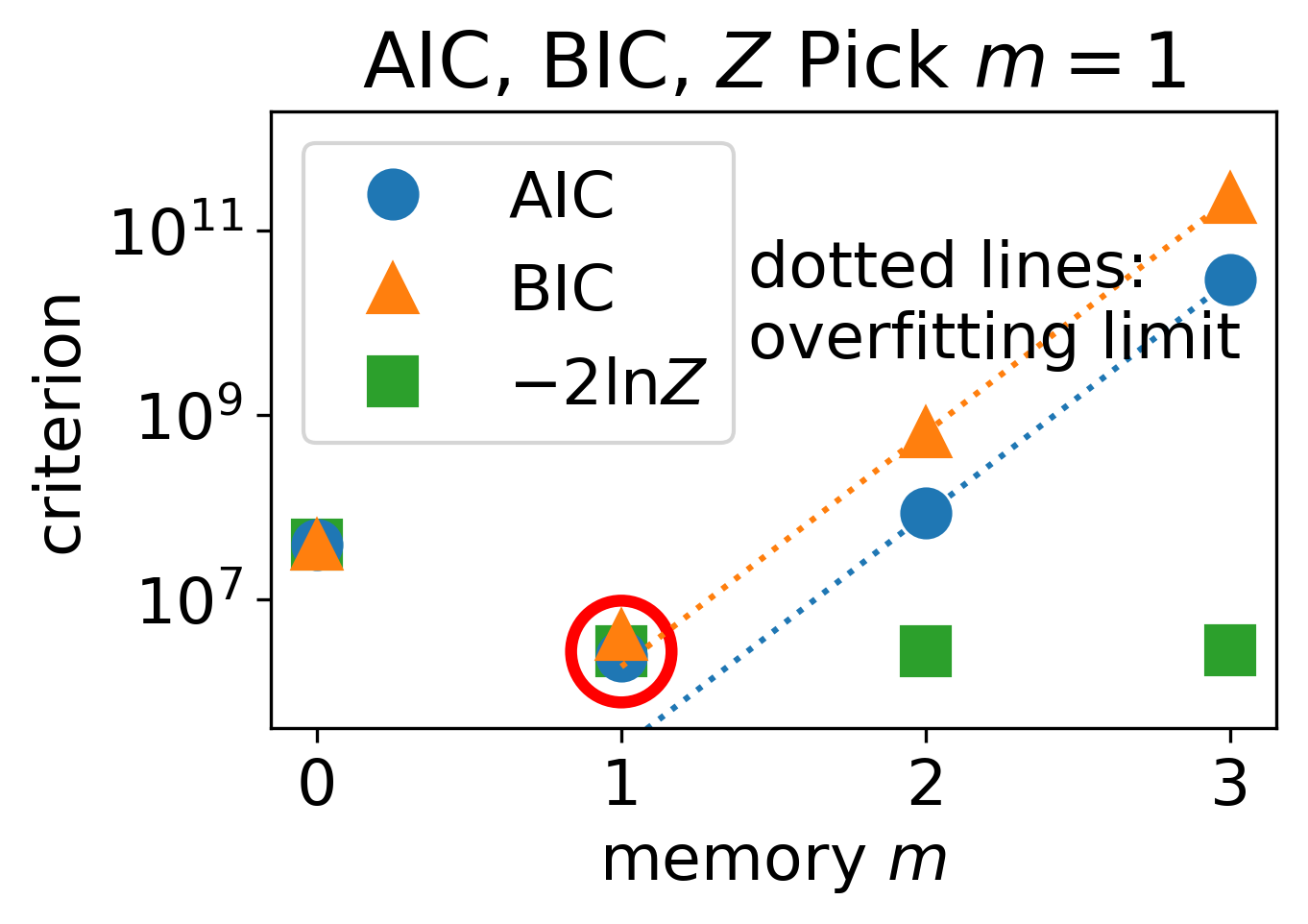}}
    \subfloat[\texttt{md5sum}]{\includegraphics[width=0.4\textwidth]{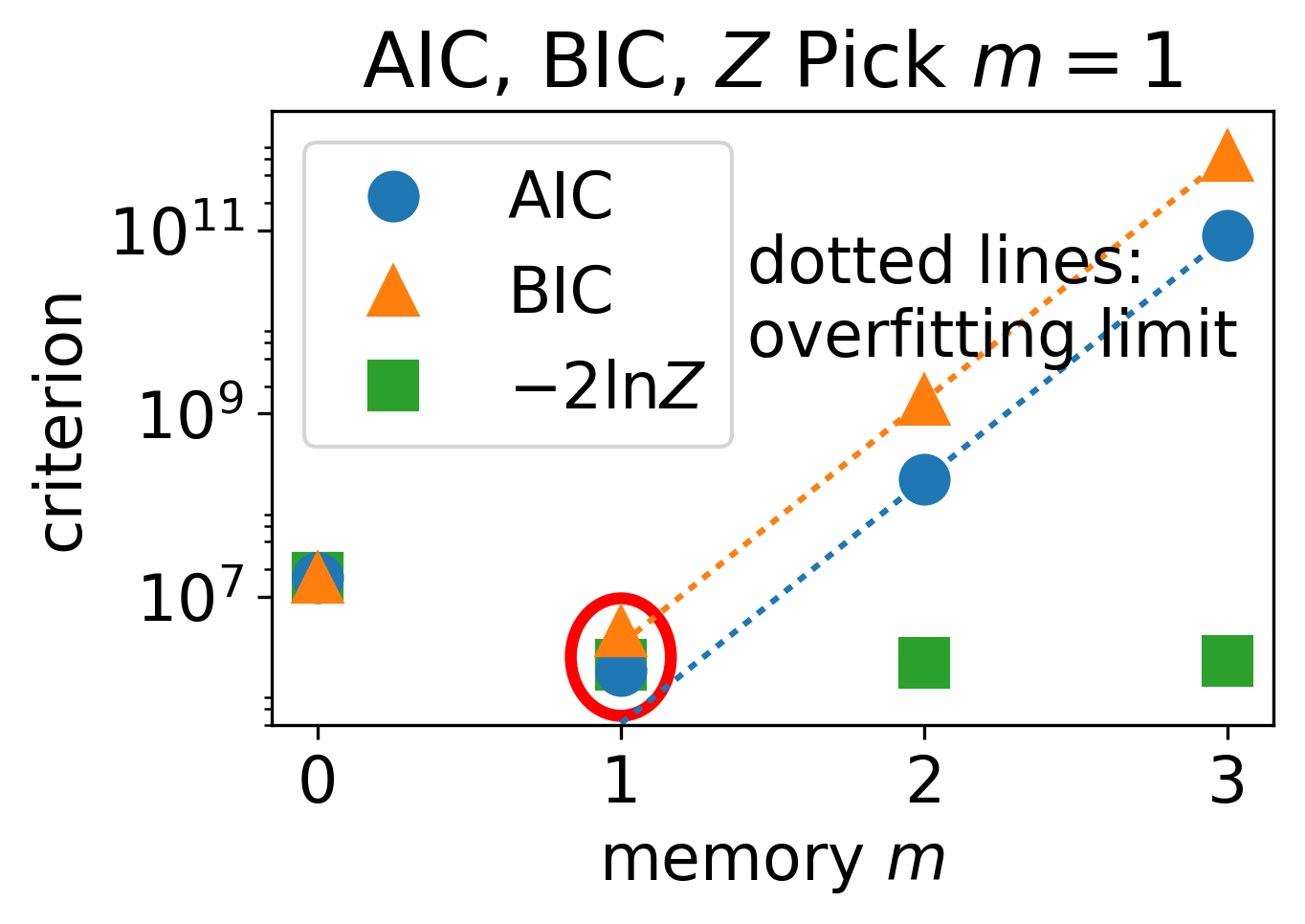}}
    \caption{All three model selection metrics select $m=1$ for both programs considered. (This includes $Z_m$, even though the increase in $-2\ln Z_m$ for $m\geq 2$ is not visible on this scale.) The $m=0$ model (states are IID) is too simple, and the $m\geq 2$ models are too complicated given the limited number of observations. The dotted lines indicate the asymptotic behavior of $AIC_m$ and $BIC_m$ once the first term in Eqs.~(\ref{eq:AIC}) and (\ref{eq:BIC}) (model complexity penalty) dominates the second term (goodness of fit). Note that the approximation in Eq.~(\ref{eq:BICLimit}) breaks down for $m\geq 2$ because $n_D < n_{\theta_m}$.}
    \label{fig:modelSelection}
\end{figure}

The three model selection metrics are shown in Fig.~\ref{fig:modelSelection}.
All three metrics prefer a first-order Markov chain, $m=1$: the $m=0$ model (states are independently and identically distributed (IID)) is too simple to explain the observed behavior, and the $m\geq 2$ models have too many parameters relative to the number of observations to be adequately constrained.

\subsubsection{Similarities may arise from shared code}
The similarities observed between \texttt{gzip} and \texttt{md5sum} in the previous sections may be explained by the code they share for initialization, finalization, input, output, and loading of shared libraries.
In order to characterize this, we constructed two simple C programs:
\begin{itemize}
	\item ``\texttt{return 0;}'' consists of ``\texttt{int main() \{return 0;\}}'' and hence only involves the code to initialize and finalize an executable.
    \item ``hello, world'' additionally exercises the code to write to the standard output.
\end{itemize}
Table~\ref{tab:overlap} summarizes the degree of overlap with these programs, where the fraction of shared instructions is defined as $(\text{\# instructions in shared functions}) / (\text{total \# instructions})$.
These results indicate that about 20\% of the functions in \texttt{gzip} and \texttt{md5sum} are there for initialization and finalization, and another 10--20\% of the functions are for output.
Also of interest is that about $50\%$ of the function in \texttt{gzip} are also in \texttt{md5sum}, though only about $16\%$ of the instructions are spent in these shared functions.
This substantial overlap illustrates why knowledge of the CFG is essential for malware detection: many benign (and malicious) programs may call the same functions but in different sequences.
Therefore, the only way to determine if the software being executed has been altered is to have knowledge of what sequences of function calls are structurally possible.

\begin{table}
	\centering
    \caption{Overlap of Functions}
    \label{tab:overlap}
    \begin{tabular}{lS[table-format=3.0]S[table-format=3.0]S[table-format=3.1]S[table-format=3.0]S[table-format=3.1]}
    	\toprule
        Program & {Functions} & \multicolumn{2}{c}{Percent shared} & \multicolumn{2}{c}{Percent shared}\\
        & & \multicolumn{2}{c}{with ``\texttt{return 0;}''} & \multicolumn{2}{c}{with ``hello, world''}\\
        \cmidrule(lr){3-4}\cmidrule(l){5-6}
         & & {Functions} & {Instructions} & {Functions} & {Instructions}\\
         \midrule
         ``\texttt{return 0;}''& 41 & 100 & 100 & 93 & 99\\
         ``hello, world'' & 82 & 46 & 5.5 & 100 & 100\\
         \texttt{gzip} & 160 & 23 & 5.5 & 28 & 5.5\\
         \texttt{md5sum} & 206 & 18 & 14 & 37 & 62\\
         \bottomrule
    \end{tabular}
\end{table}

\section{Summary and Future Work}
\label{sec:summary}
In this paper we described a general framework for detecting unintended and possibly malicious code running on processors via measurements of unintended RF emissions on a sensor separated by an air-gap. In terms of analysis, we primarily focused on the ability to track program progression at fine granularity as it traverses its control flow graph. This is a prerequisite to addressing the issue of cyber intrusion detection within this context. The approach requires principled understanding of the CFG as observed via the measurements and the fundamental limits on the ability to track the program's execution. Moreover we developed and described a principled approach for modifying the CFG of the program to dramatically enhance this tracking capability such that the program functionality is preserved and with minimal impact on overall latency of its execution. Finally, we reported some results on the extraction of CFGs and their properties, as well as the model framework describing tracking execution, for some real-world software programs.

Significant further enhancements to these methods are possible and are being investigated by our team. These include the following: 

\begin{itemize}

\item Developing a general taxonomy of colored graph models and their mitigation approaches; 
\item Applying CFG modification and tracking techniques to real software programs;
\item Understanding the trade-offs between tracking granularity and limits on tracking performance, adding new constraints to the program modification to minimize various metrics related to undesired impact on program operation;
\item Leveraging the above techniques for cyber intrusion detection (using RF emissions) for a variety of relevant use cases;
\item Exploring use cases involving more complex multi-threaded and otherwise time-shared computing environments;
\item Additional analyses of fundamental mathematical descriptions of CFG tracking in more complicated environments.
\end{itemize}

\section*{Acknowledgments}
Distribution Statement ``A'' (Approved for Public Release, Distribution unlimited).

The research effort depicted was sponsored by the Air Force Research Laboratory (AFRL) and the Defense Advanced Research Projects Agency (DARPA) under the Leveraging the Analog Domain for Security (LADS) program under contract number FA8650-16-C-7622. In particular, we thank Dr.\ Angelos Keromytis, the DARPA program manager of LADS, for his encouragement and support throughout the program. The views, opinions and/or findings expressed are those of the author and should not be interpreted as representing the official views or policies of the Department of Defense or the U.S.\ Government.

\bibliography{lads}
\bibliographystyle{plain}

\end{document}